\documentclass[10pt,letterpaper]{article}
\usepackage[top=0.85in,left=2in,footskip=0.75in,marginparwidth=2in]{geometry}

\usepackage[utf8]{inputenc}

\usepackage{cite}

\usepackage{nameref,hyperref}

\usepackage{microtype}
\DisableLigatures[f]{encoding = *, family = * }

\raggedright
\usepackage{changepage}

\usepackage[aboveskip=1pt,labelfont=bf,labelsep=period,singlelinecheck=off]{caption}

\makeatletter
\renewcommand{\@biblabel}[1]{\quad#1.}
\makeatother

\usepackage{lastpage,fancyhdr,graphicx}
\usepackage{epstopdf}
\fancyheadoffset[L]{2.25in}
\fancyfootoffset[L]{2.25in}

\usepackage{color}

\definecolor{Gray}{gray}{.25}

\usepackage{graphicx}

\usepackage{sidecap}

\usepackage{wrapfig}
\usepackage[pscoord]{eso-pic}
\usepackage[fulladjust]{marginnote}
\reversemarginpar

\begin{document}
\vspace*{0.35in}

\begin{flushleft}
{\Large
\textbf\newline{Guidelines for benchmarking of optimization approaches for fitting mathematical models}
}
\newline
\\
Clemens Kreutz\textsuperscript{1,2*}
\\
\bigskip
\bf{1} Institute of Medical Biometry and Statistics, Faculty of Medicine and Medical Center, University of Freiburg, Stefan-Meier-Str. 26, 79104 Freiburg, Germany
\\
\bf{2} CIBSS – Centre for Integrative Biological Signalling Studies, University of Freiburg, 79104 Freiburg, Germany
\\
\bigskip
* ckreutz@imbi.uni-freiburg.de

\end{flushleft}

\section*{Abstract}
Insufficient performance of optimization approaches for fitting of mathematical models is still a major bottleneck in systems biology.
In this manuscript, the reasons and methodological challenges are summarized as well as their impact in benchmark studies.
Important aspects for increasing evidence of outcomes of benchmark analyses are discussed.
Based on general guidelines for benchmarking in computational biology,
a collection of tailored guidelines is presented for performing informative and unbiased benchmarking of optimization-based fitting approaches.
Comprehensive benchmark studies based on these recommendations are urgently required for establishing of a robust and reliable methodology for the systems biology community.

\section*{Introduction}
A broad range of mathematical models are applied in systems biology.
Depending on the questions of interest and on the amount of available data, the type of models and the level of detail vary.
Most frequently, \emph{ordinary differential equation models (ODEs)} are applied because they enable a non-discretized description of the dynamics of a system and allow for quantitative evaluation of experimental data including statistical interpretations in terms of confidence and significance.
In the BioModels Database \cite{li2010biomodels}, currently 83\% of all models which are uniquely assigned to a modelling approach are ODE models.
In this manuscript, we focus on optimization-based fitting of these models although many aspects are general, and thus also relevant for other modelling types and approaches.

Abundances of compounds, the strengths and velocities of biochemical interactions are typically context-dependent, i.e.~vary between species, tissues and cell types. 
Thus, they are represented as unknown parameters in mathematical models.
Application-specific calibration of these models is therefore required which corresponds to estimation of these unknown parameters based on experimental data.

In most cases, parameter estimation is performed by optimization of a suitable \emph{objective function}
like minimization of the sum of squared residuals for \emph{least-squares} estimation, or maximization of the likelihood for \emph{maximum likelihood estimation} \cite{ashyraliyev2009systems}.
Although parameter estimation is a central task of modelling, lack of reliable computational approaches for fitting is still a bottleneck in systems biology.
The absence of high-performing software implementations seems a major reason, why ODE-based modelling is not yet a routinely applied computational approach for analyzing experimental data.

The importance of proper designs for benchmark studies in computational biology has been discussed in several publications \cite{boulesteix2018necessity, ioannidis2018meta, Kreutz16}. 
General guidelines have been provided recently for computational analysis of omics data \cite{mangul2019systematic}, for multiple alignment of protein sequences \cite{aniba2010issues}, for supervised classification methods \cite{boulesteix2013plea} as well as for periodic scientific benchmarking \cite{capella2017lessons} and for general studies in computational biology \cite{boulesteix2015ten,peters2018putting,weber2019essential}.
In this manuscript, these aspects are discussed in the context of benchmarking optimization approaches for fitting mathematical models in systems biology.

\section*{Why is reliable fitting challenging?}
A major characteristic of the mathematical models applied in systems biology is the intention to
mirror the biological process of interest because this facilitates enhanced possibilities of interpretations and understanding.
For this purpose, molecular compounds like proteins and spatial compartments like cells are defined as model components 
and regulatory interactions between the considered compounds are translated as interactions into the model.
In contrast to \emph{phenomenological} models which describe empirical relationships in an abstract and simplified manner,
the complexity of these so-called \emph{mechanistic} models is dictated by the complexity of the investigated biological process.

In systems biology, the amount of available experimental data for parameter estimation is always limited.
In such settings, multiple parameter combinations can yield to the same model response for experimentally investigated conditions.
A measured steady-state concentration, as an example, might only provide information about the ratio $k_{prod}/k_{deg}$ of production and degradation rates.
Because, different combinations of the individual parameters $k_{prod}$ and $k_{deg}$ yield the same steady state, all combinations with the same ratio fit the data equally well.
This characteristic has been termed \emph{non-identifiability}.
Because of non-identifiability, ill-conditioning and entirely flat manifolds are common features of systems biological models.

The following typical attributes of mechanistic models in systems biology raise methodological challenges for model fitting:
\begin{enumerate} 
\item The models are large in terms of number of parameters. Thus, optimization has to be performed in high-dimen\-sion\-al spaces.
\item The objective functions used for fitting depend (strongly) non-linear on the model parameters. 
Therefore, numerical optimization is intricate and local optima can exist.
\item Evaluation of the objective function comprises numerical integration of the ODEs which is computationally demanding and only feasible with limited numerical accuracy.
\item Integration of the ODEs cannot be performed analytically. Therefore, exact mathematical calculations are infeasible.
\item Derivatives of the objective function have limited numerical accuracy and cannot be calculated naively.
\item Parameter values vary over several orders of magnitudes and usually only a limited amount of prior knowledge is available. Therefore, it is difficult to specify priors, initial guesses, and/or bounds.
\item Optimization has to cope with constraints like upper and lower bounds and with non-identifiability, i.e.~with ill-conditioning and entirely flat manifolds.
\end{enumerate}

In addition, discontinuities of external inputs (so-called \emph{events}) \cite{frohlich2016parameter} might occur which has to be handled properly.
Sometimes, steady state initial values have to be calculated numerically which is another source of numerical inaccuracy \cite{fiedler2016tailored}.

\section*{Existing approaches and benchmark studies}
\begin{figure}[h!]
	\centering
  \includegraphics[width = 0.7 \linewidth]{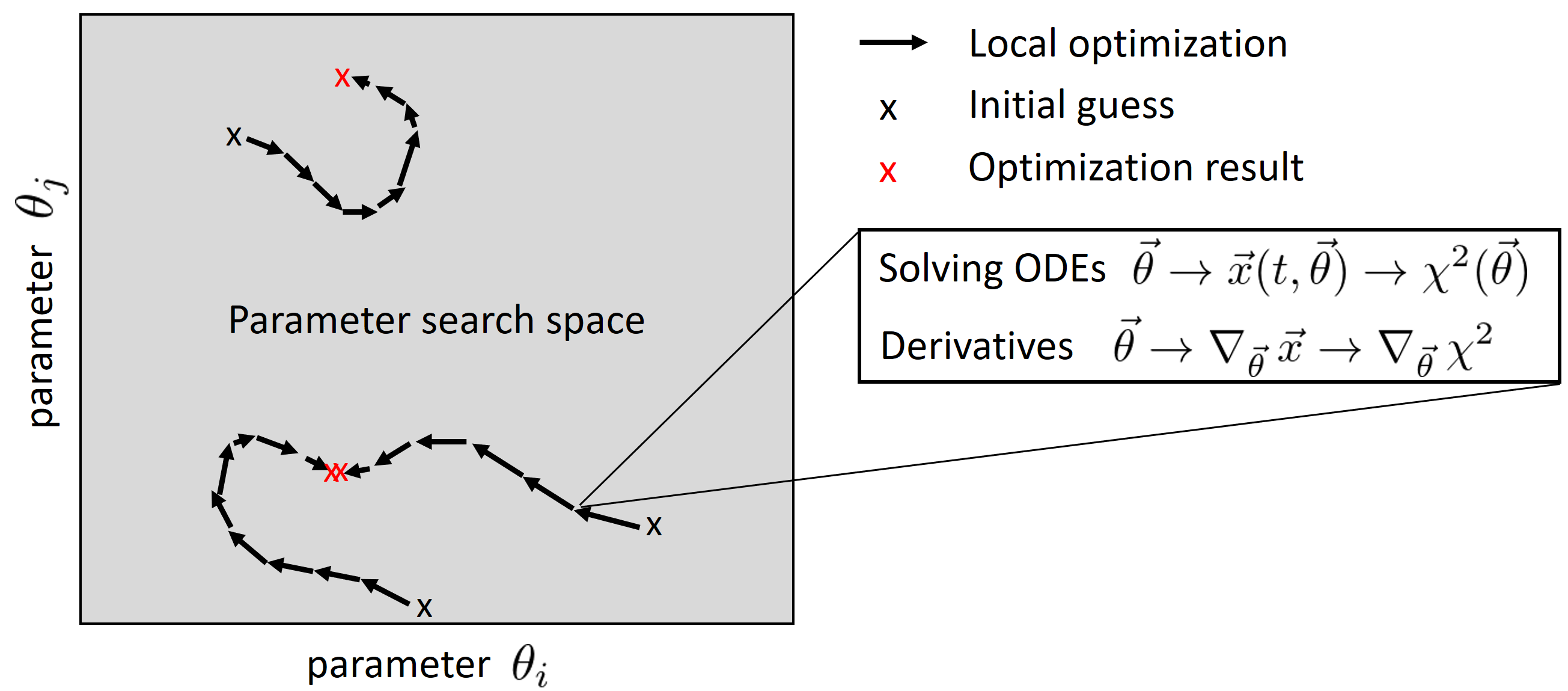}
  \caption{{\small {\bf Tasks to be accomplished for fitting ODE models.}
  Fitting of ODE models requires several tasks. Most approaches combine both, global and local search strategies.
  A prominent global search strategy is random drawing of multiple initial guesses and performing local optimization for each starting point.
  In each optimization step, the ODEs have to be solved for evaluation of the objective function $\chi^2(\theta)$.
  Moreover, an incremental improvement strategy is required for suggesting a new parameter vector for the next iteration step.
  This task is usually performed based on derivatives or approximations thereof.   \label{fig:tasks}}}
\end{figure}

For fitting of ODE models several computational tasks have to be combined as indicated in Figure \ref{fig:tasks}.
For these tasks, a multitude of methods have been published (e.g.~as summarized in \cite{ashyraliyev2009systems,frohlich2019scalable,sun2012parameter}).
For a given parameter vector, the ODEs have to be solved by numerical ODE integration methods in order to evaluate the objective function.
Moreover, iterative optimization requires a strategy for suggesting the next trial parameters in each optimization step.
This requires calculation or approximation of derivatives or alternative approaches for incremental improvement.
Such iterative local optimization approaches are typically deterministic and are usually combined with a stochastic global search strategy in order to enable convergence to the global optimum.

Currently, the existing benchmark studies provide only a heterogeneous, fragmentary and partly inconsistent picture
about the performance and applicability of fitting techniques and how these different numerical tasks have to be efficiently combined.

An early publication in systems biology found superior performance of so-called \emph{multiple shooting} \cite{peifer2007parameter} for local optimization.
This method, however, is difficult or even impossible to apply for partly observed systems with complex observation functions.
It requires a custom implementation of ODE integration method and combination with global search strategies was not considered.
Moreover, there are no publicly available implementations.
Consequently, this methodology disappeared from the systems biology field during recent years.

Multi-start deterministic optimization showed superior performance in the \emph{Dialogue for Reverse Engineering Assessments and Methods (DREAM)} benchmark challenges about parameter estimation \cite{steiert2012experimental} and network reconstruction \cite{meyer2014network}.
Here, a \emph{trust-region} and gradient based deterministic nonlinear least squares optimization approach \cite{coleman1996interior} has been utilized as local optimization strategy and global search was performed by utilizing multiple runs with random initial guesses.
This approach is implemented in the \emph{Data2Dynamics} modelling framework \cite{Raue2015} and has been shown as superior to other approaches in several studies \cite{raue2013lessons, Wieland16, Schweiger17}.

In \cite{raue2013lessons} it has been shown that deterministic gradient-based optimization is superior to a batch of hybrid and pure stochastic algorithms.
In contrast, \cite{egea2009dynamic, egea2010evolutionary, gabor2015robust, moles2003parameter} found superior performance of stochastic optimization methods.  
In \cite{villaverde2018benchmarking} it was confirmed that multi-start gradient-based local optimization is often a successful strategy but, on average, a better performance could be obtained with a hybrid metaheuristic combining deterministic gradient-based optimization with a global scatter search metaheuristic \cite{egea2009dynamic}. 

In \cite{raue2013lessons,frohlich2017scalable}, it was shown that derivative calculation based on finite difference is inappropriate.
This outcome is partly questioned in \cite{degasperi2017performance}, at least if optimization is performed on a normalized data scale.
\cite{frohlich2017scalable} showed that so-called \emph{adjoint-sensitivities} are computationally most efficient for derivative calculations for large models.

It has been repeatedly claimed that parameters are preferably optimized on the log-scale \cite{Kreutz16, villaverde2018benchmarking, Hass2019}.
Nevertheless, optimization of model parameters is still frequently performed at the linear scale in applications and even in benchmark studies like \cite{degasperi2017performance}.

In addition to neutral publications which focus primarily on benchmarking, there are additional papers which focus on introducing a new approach and compare the performance of alternative approaches in less comprehensive manner.
Despite this rather large amount of studies, there is currently
no consensus and there are no clear rules in the systems biology community about proper selection of approaches for fitting.
This shows that up to now, benchmark studies do not provide convincing evidence
which reveals the requirement for improved benchmark analyses.

\section*{Pitfalls of benchmark studies}
\subsection*{P1: Unrealistic setup}
Realistic benchmark studies have to utilize restrict to the identical setting and to the same amount of information which is available in an application setting.
Simulated data deviate from this ideal setting because real data in molecular biology typically contain non-trivial correlations, artifacts or systematic errors which is usually not considered for simulated data sets. 
Moreover, for rejecting incomplete model structures, fitting of experimental data must also work for wrong models.
Since simulated data is typically generated with the same model structure used for fitting, there is usually no mismatch between model and data, and thus optimization is only evaluated for settings where a correct model structure is available.
In order to not rely on such critical assumptions, it is strongly preferable to assess fitting approaches based on real experimental data.

In real applications, there is no information available for tuning of configuration parameters
like integration tolerances or thresholds defining termination criteria for iterative optimization.
Therefore, default configuration parameters have to be utilized or they have to be defined by consistency checks.
In order to prevent unrealistic performance assessment, one has to strictly avoid tuning of configuration parameters
based on performance criteria which are not available in practice. 
Instead, it should be prespecified before evaluation of the methods, how configuration parameters will be chosen.
Moreover, the process of tuning these parameter has to be counted as additional runtime.

\subsection*{P2: Ignoring cofactors}
\begin{table}[h!]
\begin{tiny}
      \begin{tabular}{cll}
        \hline
        Abbrev. & Cofactor   & Typical possible choices \\ \hline
        C1   &  Application problem & Model equations and the data set(s) \\
        C2   &  Primary performance criteria & Convergence per computation time, iteration steps \\
        C3   &  Secondary performance criteria & Documentation, user-friendlyness, code quality \\
        C4   &  Parameter scale 	& Linear vs. log-scale \\
        C5	 & Global search strategy  & Multiple initial guesses, scatter search algorithms, ...\\
        C6   &  Initial guess 			& Fixed vs. random \\
        C7   &  Initial guess distribution & Normal vs. uniform vs. latin-hypercube \\
        C8   &  Parameter constraints & Upper and lower bounds \\ 
        C9   &  Prior knowledge & None vs. (weakly) informative priors \\
        C10   &  ODE integration implementation & SUNDIALS, Matlab, R, ... \\ 
        C11   &  ODE integration algorithm & Stiff vs. non-stiff approaches, Adams-Moulton vs. BDF, ... \\ 
        C12   &  Integration accuracy & ODE integrator tolerances \\ 
        C13   &  Derivative calculation & Finite differences, sensitivity equations, adjoint sensitivities  \\
        C14   &  Stopping rule & Optimization termination criteria \\
        C15   &  Handling of non-converging integration & Termination of optimization vs. infinite loss \\ 
        C16   &  Algorithm-specific configurations & Cross-over rate, annealing temperature, number of particles, ... \\
        
        \hline
      \end{tabular}
\end{tiny}
\caption{\small {\bf Cofactors.} The performance of an optimization approaches depends on many decisions and configurations C1-C16.
For comparison of several approaches, these attributes appear as cofactors.
Performance benefits for individual choices do not necessarily indicate a general advantage
because benefits might originate from the chosen configurations. \label{tab:cofactors}
}
\end{table}
  \begin{figure}[h!]
	\centering
  \includegraphics[width = 0.6 \linewidth]{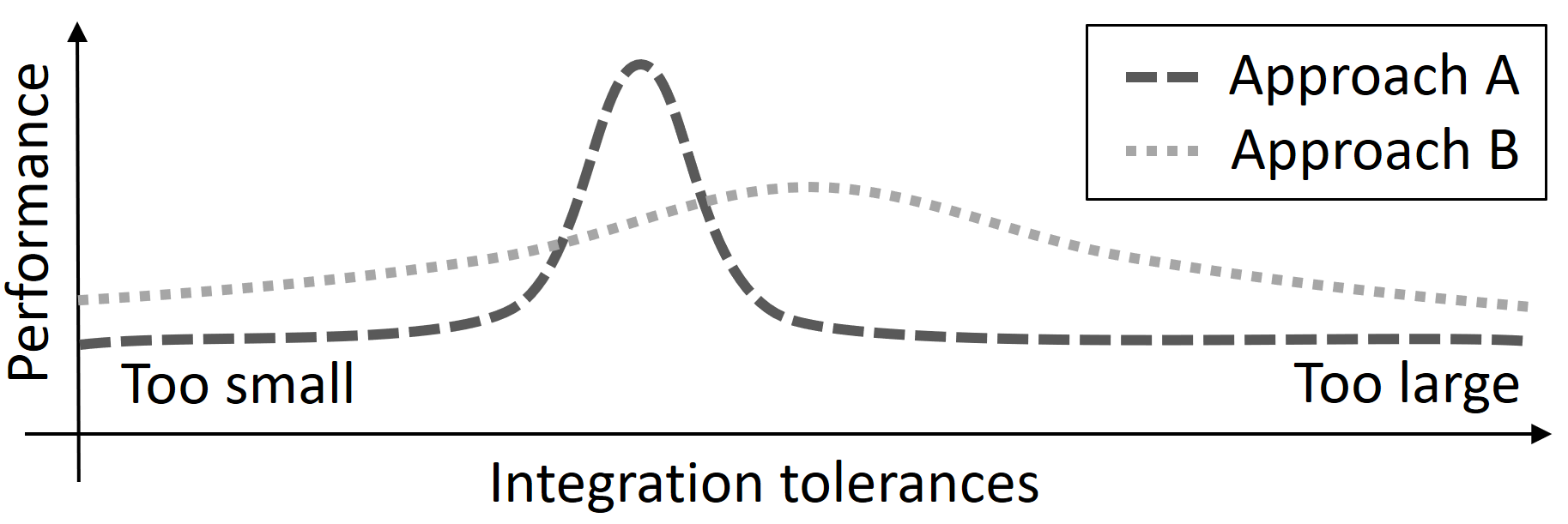}
  \caption{\small{\bf Impact of configuration parameters.}
  Multiple configurations can have an impact on the performance and performance benefits. 
  In this illustration example, two optimization approaches have different sensitivity with respect to the choice of tolerances controlling the numerical error of ODE integration. Moreover, both approaches have different optimal choices for this tuning parameter.
  For most integration tolerances, approach B is superior. However, the overall best performance has approach A for optimally chosen tolerances.
  This illustration example highlights the importance of evaluation of configuration parameters for drawing valid conclusions.
         \label{fig:multi}}
      \end{figure}
A lot of decisions have to be made for benchmarking of optimization approaches.
First, application problems have to be selected.
Then, a strategy for combining global and local search has to be specified and an ODE integration method has to be selected.
Moreover, parameter bounds and parameter scales (linear vs. logarithmic) have to be defined, ODE integration algorithms and tolerances have to be specified as well as stopping criteria for iterative optimization.
Each of these decisions or \emph{cofactors} can have a remarkable impact on the outcome of a benchmark study.

The severe problem is that these factors affect the performance of individual optimization approaches.
In the example illustrated in Figure \ref{fig:multi}, approach A requires properly tuned tolerances controlling the accuracy of ODE integration. 
If properly chosen, the approach is superior.
In contrast, approach B is less sensitive to these tolerances and outperforms A for most choices although the approach can never reach the maximal performance which is possible for approach A.

Thinking in a multi-factorial terminology is not yet common for benchmark studies, but well-established e.g.~for clinical studies
where patient-specific covariables like gender, age, smoking, etc.~occur as cofactors.
Table \ref{tab:cofactors} summarizes 16 typical cofactors for optimization benchmark studies and illustrates that many decisions have to be made to implement an optimization-based parameter estimation approach.

The impact of cofactors can be minimized by proper study designs.
Moreover, one can estimate the impact of cofactors on optimization performance by multivariate analysis of benchmarking outcomes. 
For the illustration example in Figure \ref{fig:multi}, one could determine the impact of the cofactor by evaluation of the whole range in the benchmark study which is also feasible in real application settings.
Treatment of cofactors in optimization benchmark studies has been discussed in \cite{Kreutz16}.

\subsection*{P3: Performing only case studies}
One of the most important cofactor is the chosen set of application problems, i.e.~the models and the data sets which are used for benchmarking.
A chosen model determines the intricacy of the optimization task in terms of dimension (number of parameters), non-linearity, ill-conditioning, local optima, amount of information provided by the data etc.~and thereby affect performance.

If only a single model is evaluated, the study has only low evidence because the behaviour of the applied approaches might be completely different for another application problem.
In terms of evidence, such a study is only a \emph{case report}.
Therefore, benchmark studies should be performed based a comprehensive set of test problems in order to draw generally valid conclusions.

\subsection*{P4: Non-convergence and local optima are hardly distinguishable}
\begin{figure}[h!]
  \includegraphics[width = \linewidth]{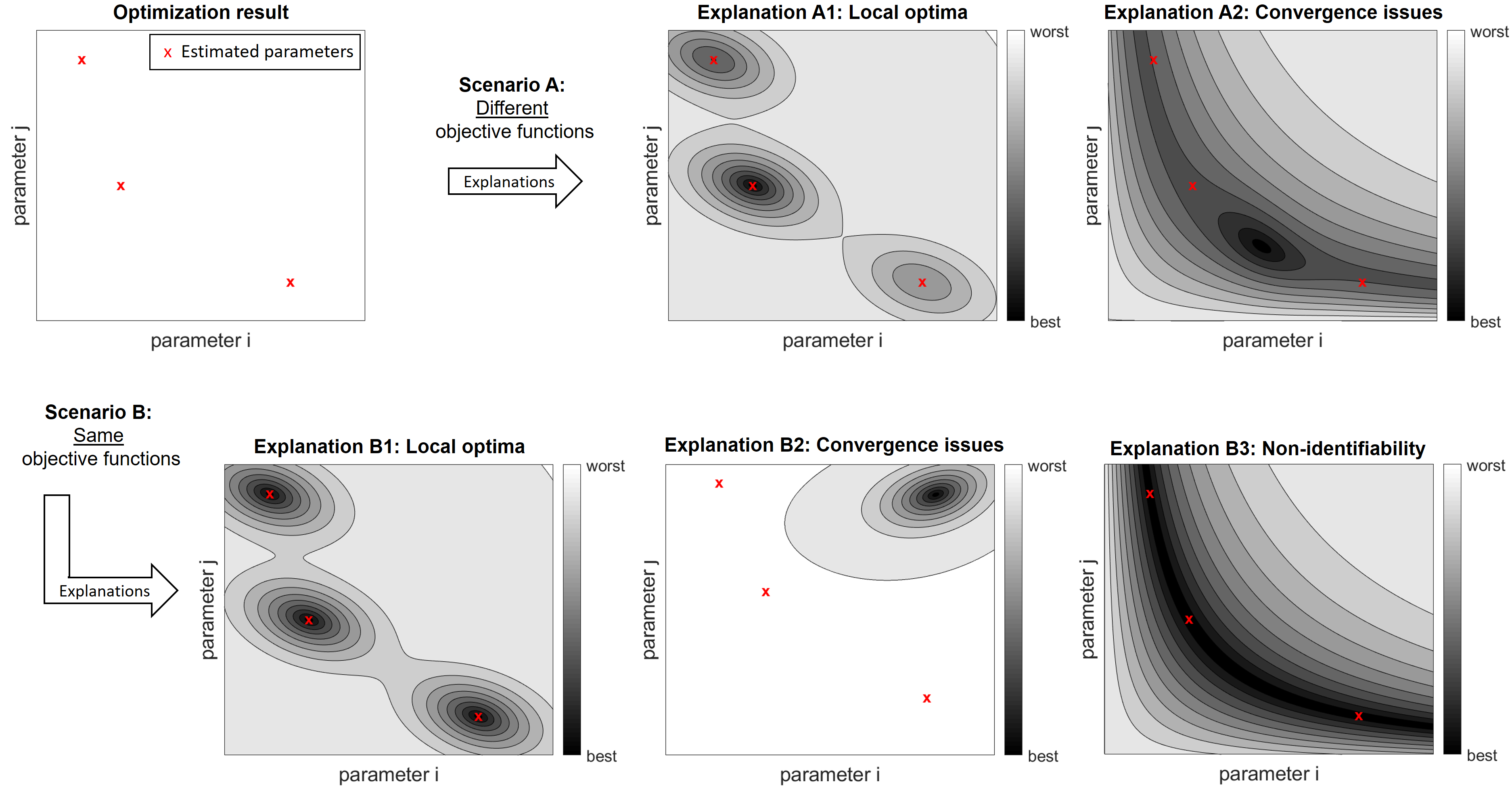}
  \caption{\small {\bf Ambiguous interpretation of optimization outcomes.} For non-trivial optimization problems, the results of independent optimization runs are typically not the same.
      The upper left panel indicates an outcome for three optimization runs, e.g.~generated with different starting points. 
      If the objective function values after optimization are different (``scenario A''), such an outcome could be explained by local optima (``explanation A1'') or by convergence problems of the optimization algorithm.
      If the same values for objective function are obtained, there might be several local optima with the same objective function (``explanation B1''), there might be a convergence problem (``B2'') or non-identifiabilities might exist (``B3''), i.e.~the estimated parameters are not uniquely specified by the data and then entire manifolds yield the same objective function.
      \label{fig:scenarios}}
\end{figure}
Sub-optimal convergence behavior is difficult to be discriminated from local optima 
since in both cases optimization terminates at distinct points in the parameter space, usually with different objective function values.
In high-dimensional spaces, it is difficult to evaluate whether a point in the parameter space is a local optimum, especially if the objective function and its derivatives can only be evaluated with limited numerical accuracy.
Existing approaches which could be applied like the profile likelihood \cite{raue2009structural}, reconstruction of flat manifolds \cite{hengl2007data}, identifiability analysis \cite{kreutz2018easy}, or methods based on a local approximation of the Hessian do not enable reliable classification in case of convergence issues and/or numerical inaccuracies.
Thus, convergence issues can easily be confused with local optima.
New methods for proving local optimality and for detection of convergence issues would be very valuable for the assessment and for future improvements of optimization approaches.

The upper left panel in Figure \ref{fig:scenarios} shows a hypothetical outcome from three optimization runs.
``Scenario A'' indicates that both, convergence problems and local optima might generate the same observations.
Even if identical values for the objective function are obtained (``scenario B''), the interpretation is ambiguous
because there are no reliable approaches to distinguish convergence issues from local optima.
Moreover, a third possible explanation in this scenario are non-identifiabilities, i.e.~multi-dimensional optimal sub-spaces.
If non-identifiabilities exist, even high-performing optimization approaches converge to distinct points in the parameter space.
For benchmark analyses, this means that measuring similarity of the outcomes of multiple optimization runs by distances in the parameter space is not reasonable.

\section*{Guidelines for benchmark studies}
Ten general guidelines for benchmark studies in whole computational biology field were recently presented in \cite{weber2019essential}.
Here, we concretize these suggestions for benchmarking of parameter optimization approaches and discuss their relevance and feasibility.
Our tailored guidelines G1-G7 and G9 correspond to the recommendations presented in \cite{weber2019essential},
guideline G8 comprises two distinct guidelines (no. 8 and 10 in \cite{weber2019essential}).
In order to provide readers a summary about the design of a benchmark study and the resulting evidence in an easily accessible and clearly structured manner, we recommend that the guidelines presented in the following are discussed in future publications in a point-by-point manner.

\subsection*{G1: Clear definition of aim and scope}
Benchmark studies which are performed to illustrate benefits of a newly presented approach are easily biased because 
they are performed to confirm merits. 
Often, application problems are utilized where existing methods have limited performance.
Consequently, it should be precisely stated, whether benchmark analyses are performed for introducing a new approach, or 
whether the goal is performing a neutral and comprehensive study based on previously published computational approaches and test cases.

Like in other fields of computational biology, 
the performance of optimization methods is context-specific, i.e.~depends on the chosen benchmark problem.
Thus, it is essential to define the scope of a study and select representative test cases according to this definition  \cite{Kreutz16}.
For mathematical modelling in the systems biology field, one could define the scope by the biological background.
Models of signalling pathways, as an example, usually describe the dynamics of activation after stimulation and show transient dynamic responses. 
In contrast, metabolic models are usually steady state descriptions.
Gene regulatory networks, on the other hand, typically have distinct rate laws because activatory and inhibitory effects are described by products of \emph{Hill equations}, instead of \emph{mass action kinetics}.
Other possibilities for defining scopes could be based on model attributes like amount of data, number of parameters, existence of events or steady state constraints, or based on dynamic characteristics like occurrence of oscillations.

\subsection*{G2: Inclusion of (all) relevant methods}
Published optimization methods are usually only available in distinct software or programming environments and require combination of several numerical tasks which do not perfectly coincide in different programming languages or software package.
As an example, the trust-region based nonlinear least-square optimization approaches implemented in Matlab and R are not identically programmed and produce different outcomes with differing performances \cite{Winkelmann16}.

Moreover, there is not yet an established standard for defining an optimization problem comprehensively including all model equations, data, measurement errors, priors, constraints, algorithms, and configuration parameters.
Thus, it is very elaborate and partly infeasible to include approaches which are previously applied in other benchmark studies.
This limitation demands for standardized data and documentation formats like \emph{COMBINE} archives \cite{bergmann2014combine}.

Nevertheless, it should be aspired to include as many approaches as possible.
For reasonable interpretation of observed performances, it absolutely is essential to implement at least one state-of-the-art approach.

\subsection*{G3: Selection of representative test cases}
Because the performance of optimization approaches strongly differ between application models, a rather large number of models is required to obtain a representative and comprehensive picture. 
However, the number of available benchmark problems is up to now strongly limited. 
Six benchmark models have been published in \cite{Villaverde2015}, however four of them only contain simulated data. 
Recently, a set of 20 benchmark problems with experimental data sets has been published \cite{Hass2019}, but it still remains difficult to perform larger studies.
Because of the small number of available test cases, it is currently only hardly feasible to define a narrow scope while still having enough test problems.

Simulating data for assessing modelling approaches
requires much more specifications than in most other fields of computational biology because realistic combinations of sampling times, observables, observation functions, error models, and experimental conditions have to be defined.
Moreover, as argued above as pitfall P1, simulation data only has limited value for benchmarking of optimization approaches.
Therefore, benchmark studies based on experimental data are currently strongly recommended.

\subsection*{G4: Appropriate configuration parameters and software versions}
Assessment of fitting approaches requires many decisions at the level of selecting approaches for the different numerical tasks as well as on the level of configuration parameters.
As discussed above (pitfall P2), configuration parameters can have a major impact on performance.
In order to guarantee fair comparisons, the strategy for their definition has to be pre-specified and should be done in an unbiased manner, i.e.~it has to be carefully evaluated that individual optimization approaches are not privileged.

According to our knowledge, existence and impact of differences between subsequent versions of the same software package 
are not yet investigated.
Nevertheless, it is essential to comprehensively describe the applied approach including software versions in order to guarantee full reproducibility.
Software tools for enhancing reproducibility of computational analyses were summarized in \cite{weber2019essential}.

\subsection*{G5: Evaluation in terms of key quantitative metrics}
Optimization is in almost all circumstances assessed by means of convergence, i.e.~in terms of probabilities or frequencies of finding local or global optima.
Although all local minima with statistically valid objective function are of interest, commonly the primary interest is the global optimum.
For proving applicability and testing performance, it is usually also sufficient to consider convergence to any kind of optimum because whether the global or a local optimum is identified is often only a question of the chosen initial guess and strongly depends on the size of the search space.
For deterministic optimization, as an example, each optimum has a region of attraction.
Hence, the frequency of finding the global optimum relative to the frequency of converging to a local optimum is mainly a question of the size of the search space and location of the optimization starting points.

Computational efforts have to be distributed among global and local search strategies, e.g.~computational efforts can be spend either for an increasing number or for increasing lengths of the individual optimization runs.
In order to balance this trade-off, it is a very reasonable strategy to assess convergence to a local/global optima by calculating the expected runtime for a single converged run as recently suggested in \cite{villaverde2018benchmarking}.
It should be kept in mind, however, that runtimes are also dictated by the computer system and especially on parallelization.
Thus, one has to ensure that rating of runtimes are fair.
Moreover, it should be investigated whether outcomes depend on the amount of parallelization, i.e.~on the number of processors, or on switching on/off parallelized code execution.

An additional issue is the definition of convergence which is typically done by thresholds for the objective function relative to the overall best known solution.
The choice of the threshold is a cofactor and its impact has to be investigated. 
Moreover, the order of magnitude has to be chosen properly, i.e.~to guarantee that only fits which are in statistical agreement with data and measurement errors are counted as successfully converged.

In most other benchmarking fields, one advantage of simulated data is knowledge about the underlying truth because then appraisal is feasible in terms of true/false positives or in terms of bias and variance.
This is different for benchmarking of optimization approaches.
On the one hand, the global solution might even for simulated data unknown because each noise realization has a different optimum.
On the other hand, outcomes from several optimization approaches can be assessed by evaluating the objective function which is feasible without any restriction for experimental data.

\subsection*{G6: Evaluation of secondary measures}
The quality of documentation, user-friendlyness, and code quality are important for diminishing the risk of incorrect application of optimization approaches. These attributes are therefore valuable as secondary measures.

Other secondary measures mentioned in the literature, are only subordinately relevant since bad convergence behavior cannot be compensated by other aspects. 
Moreover, traditional trade-offs like between precision and recall or between bias and variance do not apply for assessing convergence of optimization algorithms.

Feasibility of adaptation of optimization algorithms and fitting approaches by users is also not a major secondary aim
because custom heuristics are not recommended from our perspective. 
Within a model calibration approach, there are many aspects and details which comprehensively have to match together. 
Therefore, algorithm development requires expert knowledge and traditional strategies for tuning optimization approaches should be exploited until sufficient convergence behavior is obtained instead of implementing a weakly tested custom solution.

\subsection*{G7: Interpretation and recommendation}
Fitting of ODE models requires several numerical tasks.
Optimization only works reliably, if all these components match together and perform sufficiently well.
Proper interpretation thus requires evaluation of the impact of all configuration options (see pitfall P2)
and the multi-factorial setup has to be accounted \cite{Kreutz16}.

Benchmark studies should provide clear recommendation rules about the selection of optimization approaches for specific scopes of application, e.g.~by deriving decision trees like ``use approach A in case X, use B otherwise''.

An advantage of benchmark studies in this field is that multiple optimization approaches can be applied subsequently
in order to optimize the objective function.
One could therefore apply multiple well-performing approaches consecutively if explicit rules for selecting single approaches cannot be derived.

\subsection*{G8: Publication and reproducible reporting of results}
Since for optimization small technical details might deteriorate performances and thereby strongly impact the outcomes of benchmark studies, it is critically essential to publish the source code for the performed analyses as well as details and versions of programming environments, operating systems, and software packages.

In order to obtain consensus within the community, it seems very promising to implement online resources for fitting.
According to \cite{weber2019essential}, we also recommend that pre-print versions of peer-reviewed articles should be published in order to speed up the distribution of results, broaden accessibility, and permitting additional feedback.

\subsection*{G9: Enabling future extensions}
It is highly valuable to design and perform benchmark analyses which allow extensions by additional optimization approaches as well as new test problems in order to permit continuous update and refinements and to avoid that studies become outdated.
This guideline presented in \cite{weber2019essential} also fully applies to benchmarking of optimization approaches.

\section*{Conclusions}
For optimization-based parameter fitting of ODE models, there are several methodological challenges. 
According to existing benchmark studies, it seems that ODE models from the systems biology field demand such a rich variety of methodological requirements that each optimization approach can fail.
Unfortunately, existing studies provide only a fragmentary and inconsistent picture about the applicability of existing approaches and there is no consensus about proper selection of optimization approaches in the systems biology field.
Thus, reliable fitting of mathematical models remains a limiting bottleneck in systems biology.

We presented pitfalls for the design, analysis and interpretation of benchmark studies.
Moreover, general guidelines from the literature were tailored to the optimization-based parameter estimation setting.
The presented pitfalls and guidelines also indicate further methodological requirements.
Standards for exchanging model analyses have to be promoted in order to permit comparisons over all software environments.
A reliable methodology for distinguishing numerical causing convergence problems from local optima has to be established.
Moreover, there is urgent need for more benchmark problems which could be partly resolved by developing methodology for simulating data in a realistic manner.

\section*{Abbreviations}
G: Guideline;
ODEs: Ordinary differential equations;
P: Pitfall

\section*{Acknowledgements}
This work was supported by the German Ministry of Education and
Research by grant EA:Sys \emph{FKZ031L0080} and by the German Research Foundation (DFG) 
under Germany's Excellence Strategy \emph{CIBSS-EXC-2189-2100249960-390939984}.

\bibliographystyle{abbrv} 
\bibliography{Kreutz19-Guidelines}      

\end{document}